\documentstyle[11pt,newpasp,twoside,epsf,psfig]{article}
\def\edcomment#1{\iffalse\marginpar{\raggedright\sl#1\/}\else\relax\fi}
\reversemarginpar

\begin{document}

\title{Studying the LSS through weak gravitational lensing maps}
 \author{Antonio C. C. Guimar\~aes}
\affil{Department of Physics, Brown University, Providence, RI 02912, USA }

\begin{abstract}
Weak gravitational lensing is a promising tool for the study of the
mass distribution in the Universe. 
Here we report some partial results\footnote{More complete and detailed results will be published elsewhere} that show how lensing
maps can be used to differentiate between cosmological models. We pay special
attention to the role of noise and smoothing. 
As an application, we use mock convergence fields constructed from N-body
simulations of the large-scale structure for three historically
important cosmological models. Various map
analyses are used, including Minkowski functionals, and their ability
to differentiate the models is calculated and discussed. 
\end{abstract}


The images of distant galaxies are tangentially stretched in relation
to mass concentrations in its light path. This weak gravitational
lensing effect can be statistically measured and enables the construction
of ``lensing maps'' (see Bartelmann \& Schneider 2001 for a review of
weak gravitational lensing). 
Here we focus on the convergence field $\kappa$, which can be calculated using
\begin{equation}
\kappa({\vec \theta}) = \frac{3H_o^2}{2} \Omega_m 
\int_0^z{g(z^\prime,z)\frac{\delta({\vec \theta},z^\prime)}{a(z^\prime)} 
dz^\prime} \; ,
\label{convergence}
\end{equation}
where $\vec \theta $ is the angular position at the map,
$g$ a geometrical weighting factor, $\delta$ the density
contrast, and $a$ the scale factor.
This map contains information about the large-scale structure of the
Universe (LSS). Our aim is to relate the characteristics of the
convergence field to the LSS (see Jain, Seljak, \& White 2000), 
and to discuss the ability of analyses of the convergence to
differentiate between cosmological models.   
Our approach is to simulate $\kappa({\vec \theta})$ and use various 
statistics to characterize it.

We used N-body simulations (Hydra code - Couchman, Thomas, \&
Pearce 1995 - with $128^3$ cold dark matter
particles in boxes of side $128 h^{-1}Mpc$) 
to create realizations of the LSS between $z=0$ and a source redshift $z=1$. 
A multiple-plane lens approximation to equation (1) was them 
used to generate 25 realizations of the convergence field. 
We considered three models: 
SCDM ($\Omega_m=1, \sigma_8=0.56$),
$\Lambda$CDM ($\Omega_m=0.3, \, \Omega_{\Lambda}=0.7, \, \sigma_8=0.99$),
and OCDM ($\Omega_m=0.3, \, \Omega_{\Lambda}=0, \, \sigma_8=0.84$). 
The indicated $\sigma_8$ represents a normalization to the cluster
abundance, and we adopt $h=0.7$. 

The generated fields had a minimum size of 9.6 degrees$^2$, in a
$1024^2$ grid. 
We used a top-hat window of radius $\theta_s$ to smooth the
convergence field, and quantified it by calculating statistical
measures.
These included the convergence field
probability distribution function (PDF),  
root mean square ($\sigma^2_{\kappa}=\left< \kappa^2 \right>$), 
skewness ($S_3=\left<\kappa^3\right>/ \sigma^4_{\kappa}$), 
angular power spectrum ($P_{\kappa}(l)=\left<|\tilde{\kappa}(l)|^2\right>$), 
and Minkowski functionals. 
Noise was included as a random Gaussian field with
variance $\sigma^2_n=\sigma^2_{\epsilon}/ 2n_g\pi\theta_s^2$ 
($\sigma^2_{\epsilon}=0.16$ was the used value for the intrinsic
ellipticity variance, and $n_g=60$ arcmin$^{-2}$ the mean galaxy
density), which was added to the pure convergence field (Van
Waerbeke 2000). 

We propose a method to quantitatively evaluate the ability of an
analysis to differentiate between two weak lensing maps.
Be $A$ and $B$ two lensing maps, and $Y$ a vector-valued analysis.
$Y_A$ and $Y_B$ are the result of applying the analysis $Y$ on the
maps $A$ and $B$. $Y_{A,B}$ is assumed to follow a
normal distribution of mean value $\bar{Y}_{A,B}$ and variance
$\sigma^2_{A,B}$.
We define the {\bf differentiation} of the two maps
under the analysis $Y(p)$ ($p$ labels the vector element) as
\begin{equation}
{\cal D}_Y(A,B) = 1 - e^{-\chi^2/2} ,
\label{differ-p}
\end{equation}
where 
$\chi^2 \equiv \sum_i{[\bar{Y}_A(p_i)-\bar{Y}_B(p_i)]^2 / 
[\sigma_A^2(p_i)+\sigma_B^2(p_i)]}$. 

The differentiation assumes a value close to zero when the maps are similar
under the considered analysis, and close to one when the maps are
very different. 
That enables one to quantitatively compare different analyses when
these are applied to the same two maps. 


\begin{figure}
\centering 
\leavevmode 
\epsfxsize=4.5in
\epsfbox{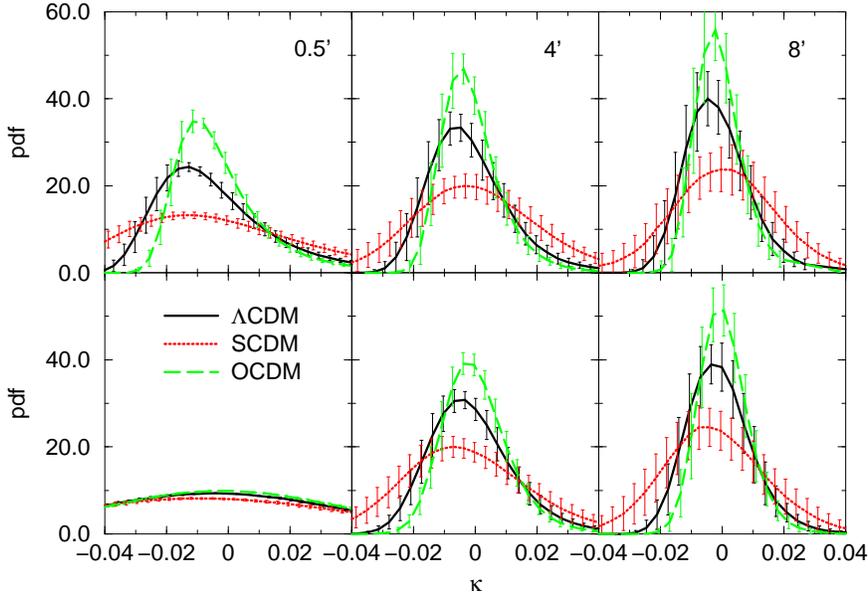}
\vspace{-0.5cm}
\caption{Probability distribution function of the convergence
  $\kappa$ for $\Lambda$CDM (solid line), SCDM (dotted line), and OCDM
  (dashed line). Upper panels are for the pure field, and bottom
  panels for the noisy field. Three smoothing scales $\theta_s$ are shown: 0.5
  arcmin (left panels), 4 arcmin (central panels), and 8 arcmin (right
  panels).}
\label{pdf-fig}
\vspace{-0.4cm}
\end{figure}

\begin{figure}
\centering 
\leavevmode 
\epsfxsize=1.8in
\epsfbox{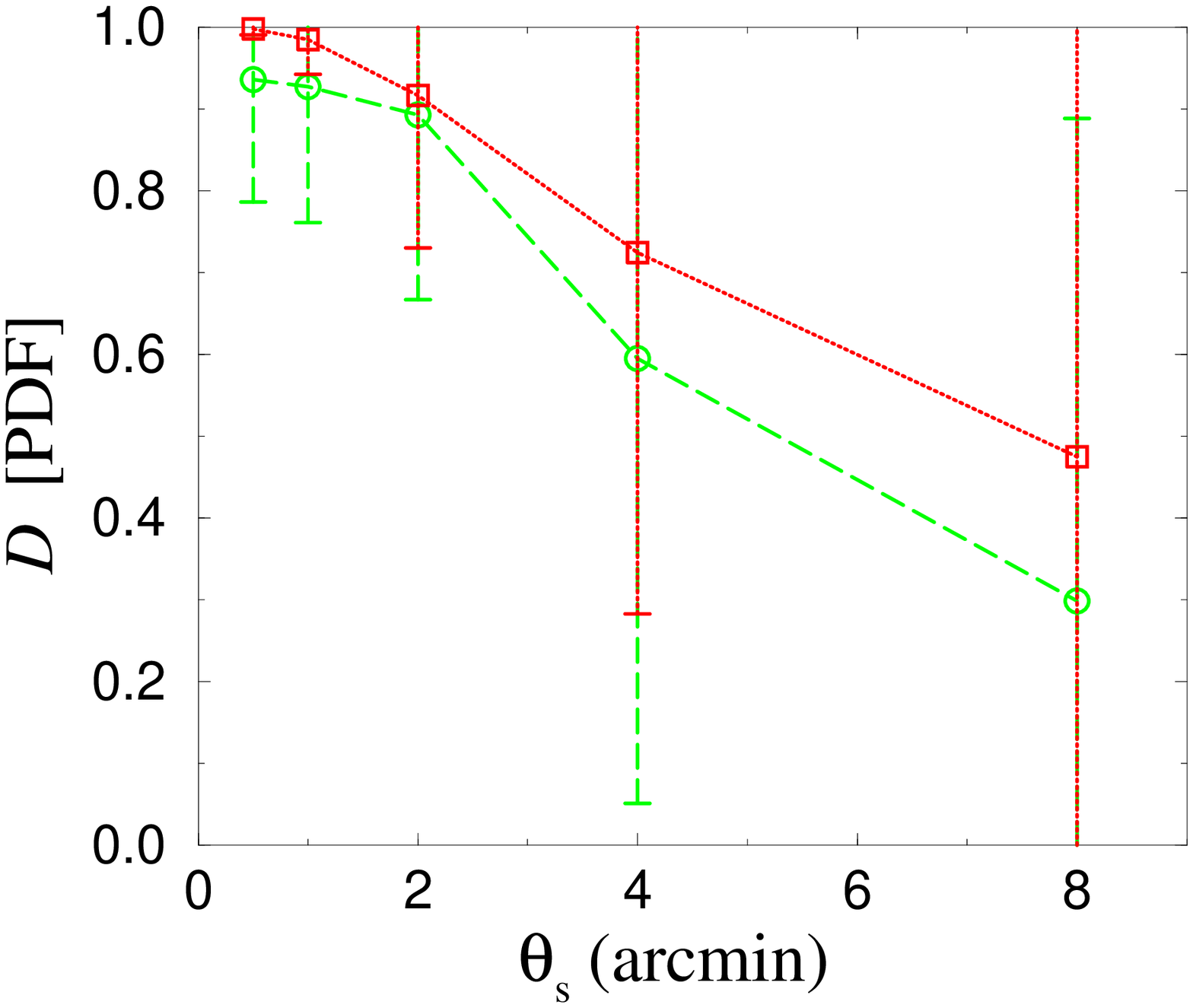}
\epsfxsize=1.8in
\epsfbox{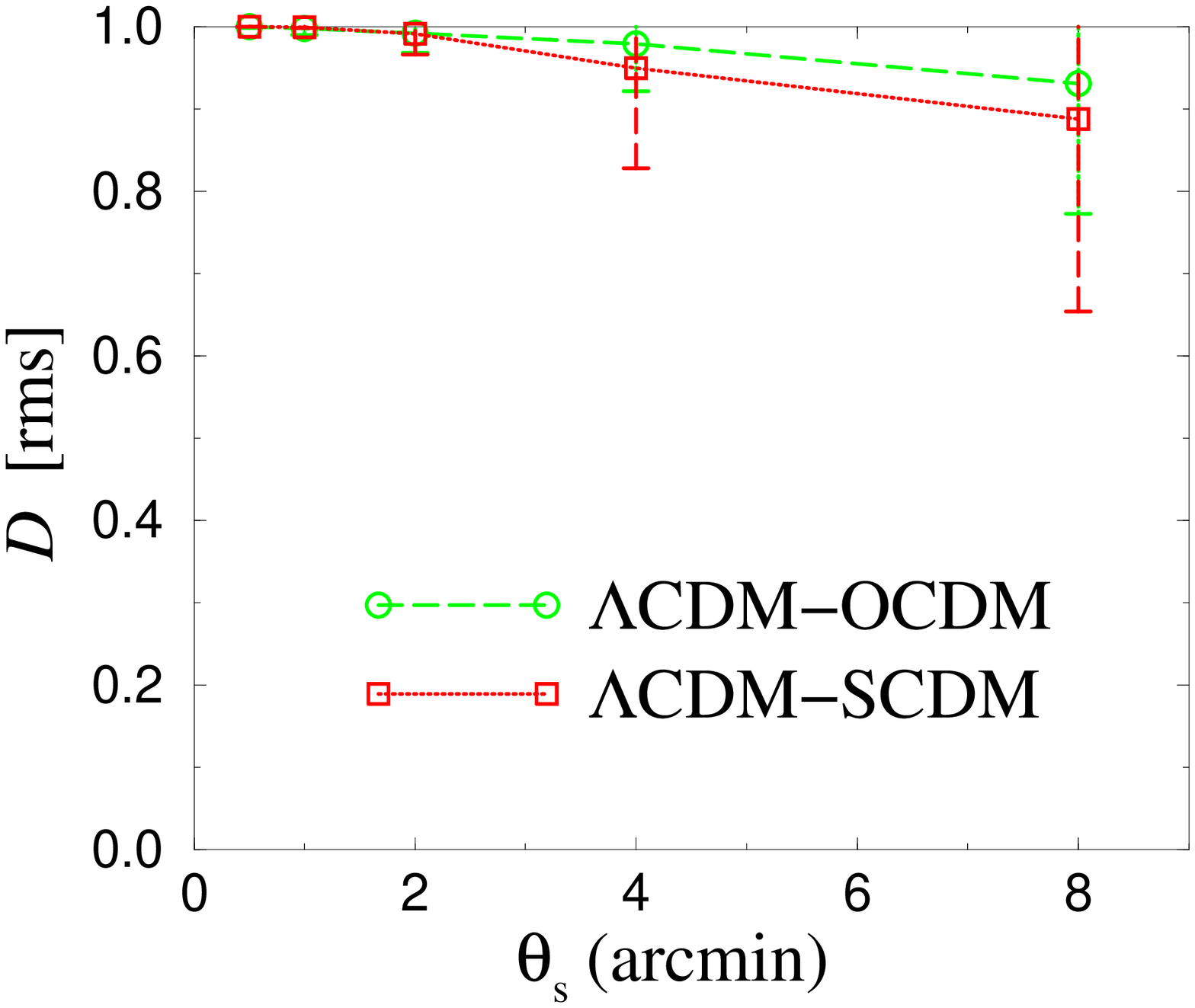}
\epsfxsize=1.8in
\epsfbox{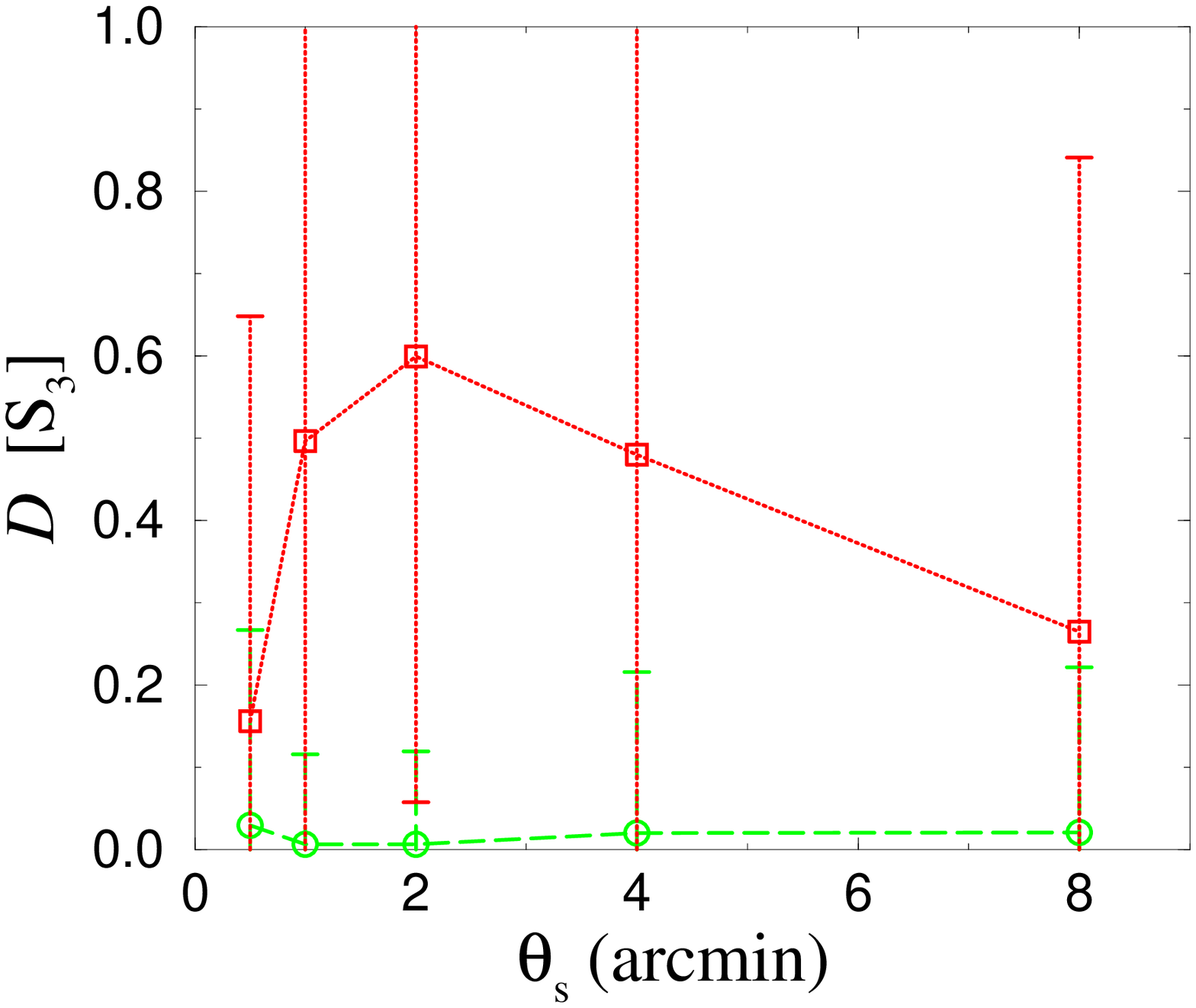}
\vspace{-0.3cm}
\caption{Differentiation between $\Lambda$CDM and SCDM (dotted line),
  and $\Lambda$CDM and OCDM (dashed line) for the analyses of the
  {\it noisy} convergence map, PDF (left panel), rms (central panel),
  and $S_3$ (right panel).} 
\label{pdf-diff}
\vspace{-0.4cm}
\end{figure}

Figure 1 shows the PDF results for the three considered models, 
which are compatible with the results of Munshi \& Jain (2000).
It illustrates the effect of smoothing and noise on the PDF
analysis of the convergence field.  
Smoothing has the effect of
reducing the root mean square (rms), or equivalently, the variance,
and skewness of $\kappa({\vec  \theta})$. 
The addition of a noise field $n({\vec  \theta})$ produces a PDF that
can be described by the convolution of the pure convergence PDF
$F_{\kappa_o}$ with the noise PDF $F_n$,
\begin{equation}
F_{\kappa}(x)=\int_{-\infty}^{+\infty}{F_{\kappa_o}(y)F_n(x-y) dy} \; ,
\end{equation}
the variance increases by the noise field variance ($\sigma_{\kappa}^2
= \sigma_{\kappa_o}^2 + \sigma_n^2$), and the skewness is reduced by a
factor, 
$S_3(\kappa)=S_3(\kappa_o) (\sigma_{\kappa_o}/\sigma_{\kappa})^4$. 

The curves for the differentiation $\cal D$ between the models are show
in Figure 2. 
As expected, the smoothing of the field tends to make
the models indistinct (decreasing $\cal D$). 
Both the PDF and the simple rms analyses are able to
distinguish the models at a significant confidence level at
low smoothing. 
However, the skewness proves to be a bad analysis for
differentiating these models (for convergence maps of the considered
size and noise level), because the obtained ${\cal D}_{S_3}$ is
close to zero between $\Lambda$CDM and OCDM, and also because it is not very
significant between $\Lambda$CDM and SCDM.


\begin{figure}
\centering 
\leavevmode
\epsfxsize=1.7in
\epsfbox{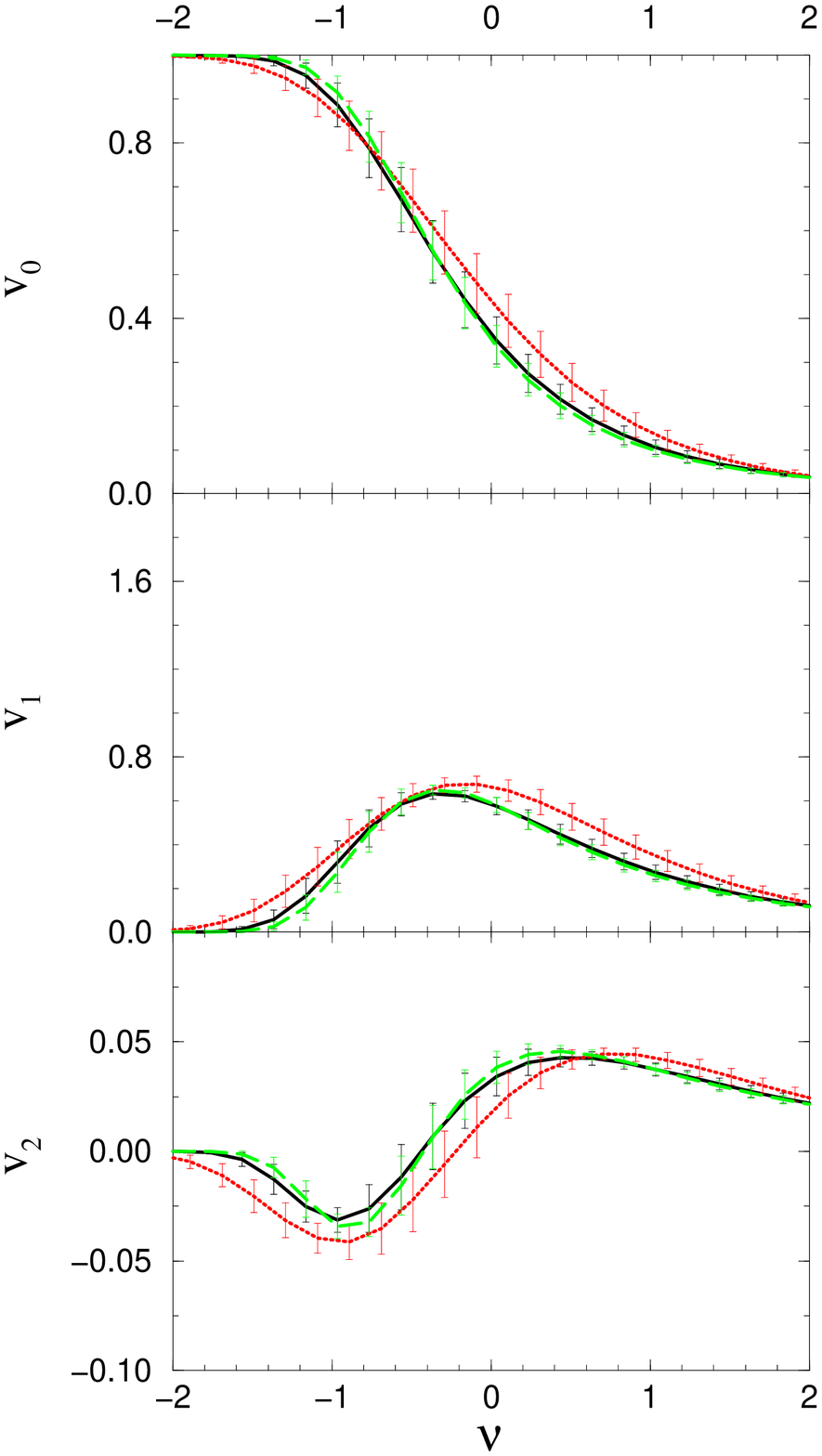}
\epsfxsize=1.7in
\epsfbox{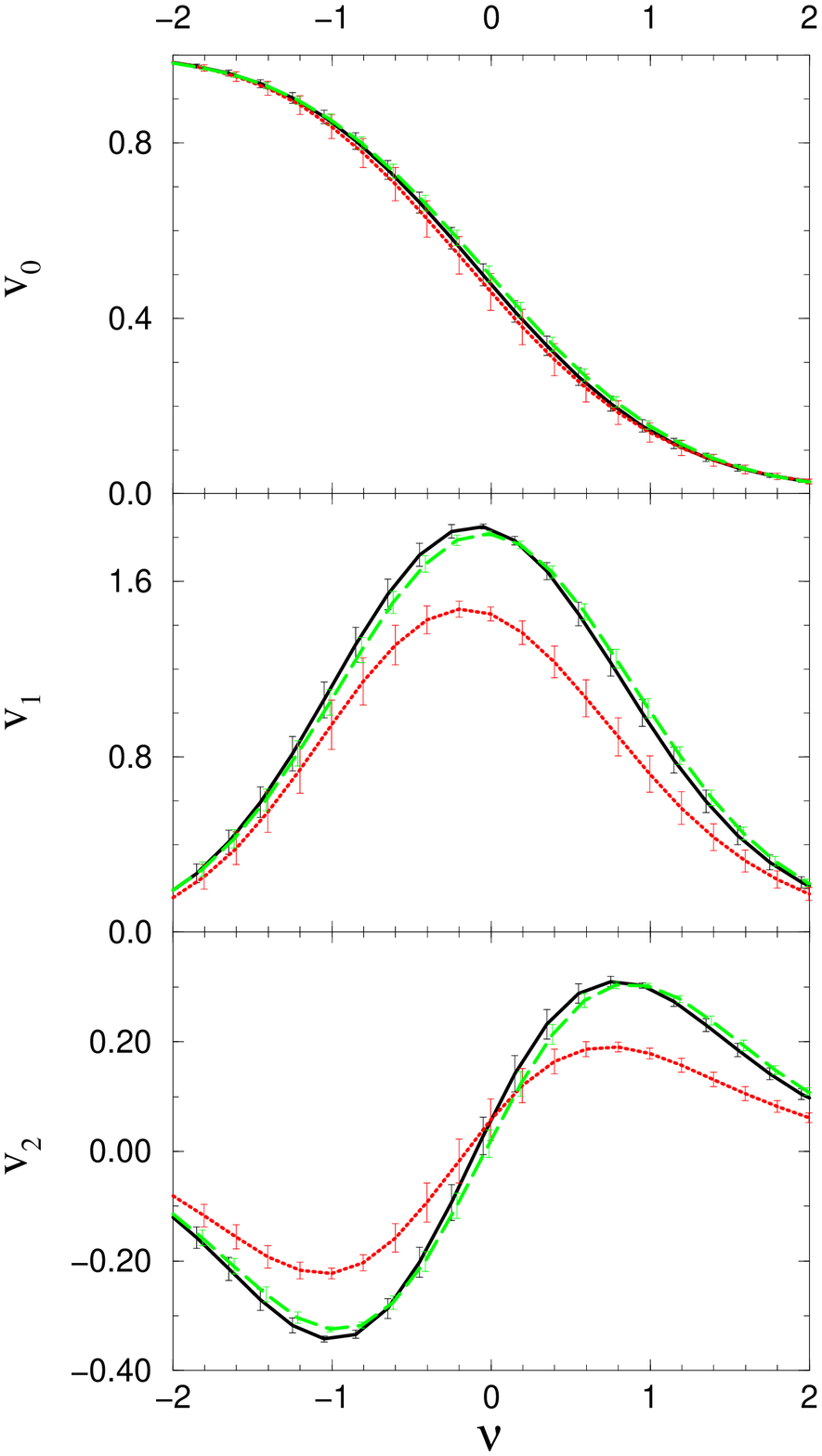}
\epsfxsize=2.5in
\epsfbox{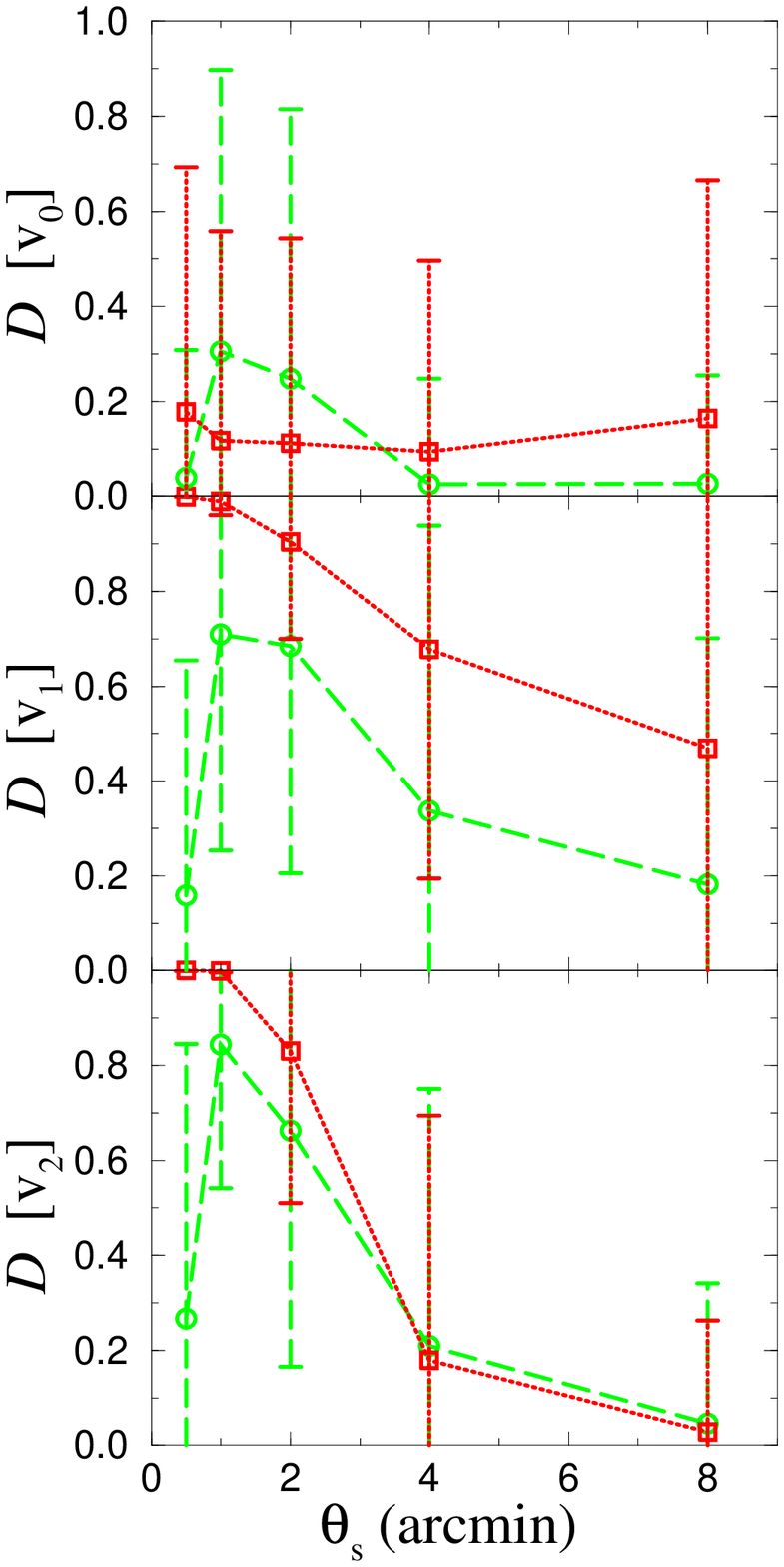}
\vspace{-0.5cm}
\caption{Minkowski functionals for the convergence field (pure at the
  left panel, and noisy at the central panel), for the three
  cosmological models considered (see Fig. 1), with $\theta_s=0.5$
  arcmin smoothing. The right panel shows
  the differentiation between two models (see Fig. 2) under the Minkowski
  functionals analysis (noise included).}
\label{mink-fig}
\vspace{-0.4cm}
\end{figure}

The addition of noise adds a power law to the power amplitude 
($P_{\kappa}=P_{\kappa_0}+P_n$). For small smoothing angular
scales this contribution is significant, but it becomes negligible
for large smoothing angular scales. However, at large smoothing angles
power on small scales (large wavenumber $l$) is suppressed - structure
is washed out by the field smoothing (figure not shown).

The use of Minkowski functionals to characterize the morphology of the
convergence field is another way to study the non-Gaussianity of
$\kappa$ (see  Winitzki \& Kosowsky 1998, Matsubara \& Jain 2001, and
Sato et al. 2001). 
Figure 3 shows some results for the Minkowski functionals analysis of
$\kappa$; the threshold $\nu\equiv\kappa/\sigma_{\kappa}$ defines a 2D
contour map, and the Minkowski functionals can be roughly described
as the fractional area inside the contours ($\nu_0$), the
boundary length ($\nu_1$), and the Euler characteristic ($\nu_2$).
Our results reveal that Minkowski functionals are very sensitive to
noise and smoothing. 
The second functional ($\nu_1$) have equal or even more discriminating 
power than the third functional ($\nu_2$), and
comparing their differentiation curves
with the one for skewness, which also reflects non-Gaussian aspects of
the convergence, their ability to discriminate between the considered
models is superior.
This differentiation is maximum for a median smoothing. At too small
smoothing scales noise dominates, and at too large smoothing scales
distinguishing features are erased.

Our results indicate that a comprehensive study of the LSS through
weak gravitational lensing maps requires the use of a set of analyses,
because different analyses reveal distinct features of the underlying
cosmic mass distribution and geometry. 
Also, the proposed quantity $\cal D$ proved to be very
useful for quantifying the differentiation between analyses of
lensing maps, and for comparing different statistical measures of lensing.

\acknowledgments{ACCG thanks Uro\v{s} Seljak for providing codes and
  valuable knowhow, the Princeton University Physics
  Department for the use of its computer facilities, and also Robert
  H. Brandenberger and Ian 
  dell'Antonio for very helpful discussions.
  The research at Brown was supported in part by the US Department of
  Energy under Contract DE-FG0291ER40688, Task A.}


\end{document}